\documentclass[twocolumn,trackchanges]{aastex63}
\usepackage{color}
\usepackage{ulem}
\usepackage{hhline}

\begin{document}

\title{Radio Study of the Pulsar Wind Nebula Powered by PSR B1706$-$44}
\author{Y. H. Liu}
\affiliation{Department of Physics, The University of Hong Kong, Pokfulam Road, Hong Kong}
\email{yihanliu@connect.hku.hk}
\author[0000-0002-5847-2612]{C.-Y.~Ng}
\affil{Department of Physics, The University of Hong Kong, Pokfulam Road, Hong Kong}
\author[0000-0003-0392-3604]{R. Dodson}
\affiliation{International Centre for Radio Astronomy Research, The University of Western Australia, 35 Stirling Hwy, Crawley,  Western Australia}
\correspondingauthor{Y. H. Liu}


\newcommand{\hand}{"The hand of God"}
\newcommand{\psr}{PSR~B1706$-$44}
\newcommand{\msh}{MSH~15$-$5\textit{2}}
\newcommand{\gname}{G343.1$-$2.3}
\newcommand{\rcw}{RCW~89}
\newcommand{\hii}{H{\sc ii}}
\newcommand{\hi}{H{\sc i}}
\newcommand{\pwn}{the PWN}
\newcommand{\Pwn}{The PWN}
\newcommand{\uv}{\textit{u-v}}
\newcommand{\perbeam}{beam$^{-1}$}
\newcommand{\JyPerBeam}{$\,{\rm Jy\,beam^{-1}}$}
\newcommand{\mJyPerBeam}{$\,{\rm mJy\,beam^{-1}}$}
\newcommand{\uJyPerBeam}{$\,\mu{\rm Jy\,beam^{-1}}$}
\newcommand{\ergPerSecond}{\,erg\ s$^{-1}$}
\newcommand{\ergcmPerSecond}{\,erg\ cm$^{-2}$ s$^{-1}$}
\newcommand{\Msun}{$M_{\odot}$}
\newcommand{\uG}{$\,\mu$G}

\newcommand{\cm}{\,cm}

\newcommand{\hour}{^{\rm h}}
\newcommand{\minute}{^{\rm m}}
\newcommand{\second}{^{\rm s}}
\newcommand{\fsecond}{\fs}
\newcommand{\degrees}{^\circ}
\newcommand{\Chandra}{\emph{Chandra}}

\newcommand{\task}[1]{\texttt{#1}}
\newcommand{\RMu}{\,rad\,{m$^{-2}$}}

\newcommand{\rd}[1]{\textcolor{red}{#1}}
\newcommand{\bl}[1]{\textcolor{blue}{#1}}
\newcommand{\gr}[1]{\textcolor{green}{#1}}
\newcommand{\yl}[1]{\textcolor{yellow}{#1}}
\newcommand{\ora}[1]{\textcolor{orange}{#1}}
\newcommand{\cy}[1]{\textcolor{cyan}{#1}}
\newcommand{\mg}[1]{\textcolor{magneta}{#1}}

\newcommand{\arcs}{\mbox{$^{\prime\prime}$}}
\newcommand{\arcm}{\mbox{$^{\prime}$}}

\begin{large}

\begin{abstract}
PSR B1706$-$44 is an energetic gamma-ray pulsar located inside
supernova remnant (SNR)\ \gname\ and it powers 
a compact pulsar wind nebula (PWN) that shows torus and jet structure in X-rays.
We present a radio study of the PWN using Australia Telescope Compact Array (ATCA) observations at 3, 6, 13, and 21\,cm.
We found an overall arc-like morphology at 3 and 6\cm, and the ``arc" shows two distinct peaks at 6\cm.
The radio emission is faint inside the X-ray PWN and only brightens beyond that.
We develop a thick torus model with Doppler boosting effect to explain
the radio PWN structure.
The model suggests a bulk flow speed of $\sim 0.2c$,
which could indicate significant deceleration of the flow 
from the X-ray emitting region.
Our polarization result reveals a highly ordered toroidal $B$-field in the PWN.
Its origin is unclear given that the supernova reverse shock should have interacted with the PWN.
At a larger scale, the 13 and 21\cm\ radio images detected
a semi-circular rim and an east-west ridge of \gname.  
We argue that the latter could possibly be a pulsar tail rather than a filament of the SNR,
as supported by the flat radio spectrum and the alignment between the magnetic field and its elongation.
\end{abstract}

\section{Introduction}

A pulsar is a compact star that is born in a supernova explosion. It emits periodic signals and has a strong surface magnetic field. Particles around a pulsar are accelerated to form relativistic wind as the pulsar spins down and loses energy constantly. The relativistic wind interacts with the ambient medium and forms a synchrotron nebula, called a pulsar wind nebula (PWN). A PWN is able to accelerate particles to very high energies, emitting synchrotron radiation from the radio to hard X-ray bands. 

In X-rays, torus-jet features are commonly detected in young PWN systems \citep{Kargaltsev2008PWNe,PWNTori_Ng2004,Ng2008PWNTori}. Theories suggest that torus structure is due to the shocked pulsar wind flowing into the equatorial region, and jets are wind in the polar regions confined by magnetic hoop stress \citep[see][]{Porth2017torijetmodel}. 
In the radio band, however, these features are rarely seen.
For instance, the Crab, 3C~58, and the PWN inside G292.0+1.8 are filled with wisps and filamentary structures \citep{Dubner2017crab,3c58wisp_Bietenholz2006, Gaensler2003PSRJ1124g292} instead;
some show arc-like structure, such as CTB~87 \citep{Kothes2020CTB87} and double-lobed morphology, such as G21.5$-$0.9, G76.9+1.0, and DA~495 \citep{Bietenholz2008J1833double-lobed,Arzoumanian2011g76.9,Kothes2008DA495}. 
Previous radio polarization observations also revealed different magnetic field configurations in young PWNe.
From theoretical works, the radial component of the magnetic field decays faster than the toroidal component
($\sim$$r^{-2}$ vs.\ $\sim$$r^{-1}$)  \citep{Porth2017torijetmodel}.
Therefore, the $B$-field beyond the termination shock should be mostly toroidal.
This, however, is not supported by observations.
Only a few cases, including Vela and Boomerang, show toroidal $B$-field \citep{Dodson2003VelaPWN,Kothes2006Boomerang},
but many others, e.g., the Crab Nebula, 3C~58, G21.5$-$0.9, and Dragonfly,
have complex or radial field structure \citep{Reich2002SNRradio,Lai2022g21.5,Jin2022Dragonfly}.
The physical cause of such diverse morphology and magnetic field structure among radio PWNe is not fully understood and a larger sample is needed for further study.

In this work, we present a new radio study of the PWN powered by the Vela-like pulsar \object{B1706$-$44}.
It is one of the few $\gamma$-ray pulsars detected in the early days with EGRET \citep{Mcadam1993B1706gamma}. It has a characteristic age $\tau_{c}=$17.1\,kyr and a spin-down power $\dot E\approx4\times10^{36}$\,erg\,s$^{-1}$. A recent study with \emph{Chandra} found that the pulsar is moving eastward with a projected velocity of around 130\,km\,s$^{-1}$ \citep{deVries_J1709_motion_2021}. The association between the pulsar and the nearby supernova remnant (SNR) \gname\ is controversial.
The remnant has a circular shell and an east-west ridge in the southern part \citep{dodson2002association}.
The pulsar is located at the tip of the ridge, near the center of the shell.
The SNR distance of $\sim3.5$\,kpc estimated from the $\Sigma$--$D$ relationship is compatible to the pulsar dispersion measure distance $d\approx 2.3\,$kpc \citep{PWNDMdistance_Cordes2002,Yao2017YMW16_DM,Mcadam1993B1706gamma}. The High Energy Stereoscopic System (H.E.S.S) detected extended TeV emission west of the pulsar, which also has some connection with the SNR \citep{G343.1-2.3TeV_HESS2011}. Besides, the pulsar powers an X-ray PWN that has compact torus and jet structure \citep{2005ApJ...631..480R}. A recent study found diffused emission around the torus and a long curved outer-jet \citep{deVries_J1709_motion_2021}.

In this study, we aim to perform high resolution radio observations of B1706 PWN for directly comparing with the compact X-ray structures in \emph{Chandra} images and better understanding the PWN magnetic properties. Previous observations have detected a radio PWN surrounding the pulsar \citep{Frail1994B1706PWNVLAradioSNRs,giacani_2001_psrB1706,dodson2002association,2005ApJ...631..480R}. However, the pulsar emission was not clearly distinguished from the PWN. Some ATCA observations excluded the pulsar emission, but few observations with high resolution and sensitivity were included. Besides, there is no previous study of the magnetic field in the B1706 PWN.

In this paper, we analyze new and archival radio observations of the PWN powered by \psr\ (hereafter B1706 PWN) and SNR \gname\ taken with the Australia Telescope Compact Array (ATCA) at 3, 6, 13, and 21\cm\ images.
We employed new observations with high resolution aiming to better study the morphology and polarization information of this PWN. 
We describe the observations and data reduction process in Section~\ref{sec2}. Section~\ref{sec3} shows the results and they are discussed in Section~\ref{sec4}. We summarize our results in Section \ref{sec5}.

\section{Observations and Data reduction}
\label{sec2}
\begin{deluxetable*}{ccccccc}[htp]
\tablecaption{ATCA observations of B1706 PWN used in this study \label{tab:1}}
\tablenum{1}
\tablehead{\colhead{Obs.\ Date} & \colhead{Array} & \colhead{Center Freq.} & \colhead{Usable Band-} & \colhead{No.\ of} & \colhead{Integration} & \colhead{Pulsar} \\
\colhead{} & \colhead{Config.} & \colhead{(MHz)} & \colhead{width (MHz)} & \colhead{Channels} & \colhead{Time (hr)} & \colhead{Binning Mode}}
\startdata
\cutinhead{\textbf{3\,cm}}
2002 Jan 06 & 750A & 8640 & 104 & 13 & 7.7 & Y\\
2002 Feb 16&1.5A&8640&104&13& 10.2&Y\\
2002 Apr 11&6A&8640&104&13& 8.8&Y\\
2003 May 18&1.5C&8384, 8640&104&13&8.7&Y\\
2003 Jun 23&750C&8384, 8640&104&13&10.2&Y\\
2003 Aug 02&6D&8384, 8640&104&13&9.8&Y\\
2005 Nov 20&1.5C&8384, 8640&104&13&4.3&Y\\
2005 Dec 27&6A&8384, 8640&104&13&4.3&Y\\
2017 Nov 03&6A&8997.5&1728&433&10.9&Y\\
2018 Jan 11&6C&8997.5&1728&433&10.3&Y\\
\cutinhead{\textbf{6\,cm}}
2002 Jan 06&750A&4800&104&13&7.7&Y\\
2002 Feb 16&1.5A&4800&104&13& 10.2&Y\\
2002 Apr 11&6A&4800&104&13& 8.8&Y\\
2017 Nov 03&6A&5497.5&1728&433&10.9&Y\\
2018 Jan 11&6C&5497.5&1728&433&10.3&Y\\
\cutinhead{\textbf{13\,cm}}
1998 May 29&750E&2496&104&13&0.8&N\\
1999 Nov 03&210&2496&104&13&19.1&N\\
\cutinhead{\textbf{21\,cm}}
1998 May 29&750E&1384&104&13&0.8&N\\
1998 Sep 15&6A&1384&104&13&3.3&N\\
1999 Nov 03&210&1384&104&13&19.1&N\\
2005 Nov 19&1.5C&1344, 1472&104&13&4.6&N\\
2005 Dec 27&6A&1344, 1472&104&13&1.7&N\\
\enddata

\end{deluxetable*}

We carried out new radio observations of B1706 PWN at 3 and 6\cm\ bands with ATCA in 6\,km array configurations on 2017 Nov 3 and 2018 Jan 11. We also analyzed archival ATCA observations taken in the 3, 6, 13, and 21\cm\ bands with various array configurations, which have previously been analyzed by \citet{dodson2002association,2005ApJ...631..480R}. All the 3 and 6\cm\ band data were taken with the pulsar binning mode, providing a high time resolution. We then only select off-pulse data to ``gate out" the pulsar emission to search for faint PWN structure in the surrounding. 

Table \ref{tab:1} lists the detailed observation parameters of all the data. The 3 and 6\cm\ observations were performed simultaneously centering at 8640\,MHz and 4800\,MHz, as well as at 8997.5\,MHz and 5497.5\,MHz. Besides, we have also selected observations in 2003 and 2005 at 8640\,MHz and 8384\,MHz. Our new observations in 2017 and 2018 were taken after the Compact Array Broadband Backend (CABB) upgrade \citep{Wilson_2011}, which increased the bandwidth from 128\,MHz to 2048\,MHz.  At 3\,cm, the pre-CABB and post-CABB integration times are 64.0\,hr and 21.2\,hr, respectively, with a total \uv\ coverage from 0.8\,k$\lambda$ to 197.4\,k$\lambda$. At 6\,cm, we have 26.7\,hr and 21.2\,hr pre- and post-CABB integration time, respectively, covering the \uv\, space from 0.85\,k$\lambda$ to 127\,k$\lambda$. The 13 and 21\cm\ datasets with good quality have a total integration time of 19.9\,hr and 29.5\,hr, respectively. The \uv\,coverage of the observations at 13\cm\ is 0.2--5.5\,k$\lambda$ and 18--37\,k$\lambda$, and in the 21\cm\ band is 0.1--30\,k$\lambda$.

We processed the data using the MIRIAD package \citep{sault1995astronomical}. We first flagged the edge channels and data affected by severe radio frequency interference, then followed the standard procedures to calibrate the flux scale, band pass, and gains. After calibration, we formed Stokes I, Q, and U images using multi-frequency synthesis. We weighted the data inversely proportion to the noise. Since the pre- and post-CABB data were taken over 15\,yr apart, we formed separated images at 6\cm\ to check for any morphological changes. The result shows no significant variability, we therefore combined all off-pulse data at each frequency for a joint analysis to boost the signal. 

We first focused on the region close to the pulsar, and generated the 3\cm\ image with the best resolution of full width half maximum (FWHM) 6.3\arcs$\times$3.5\arcs. The resulting map has root mean square (rms) noise of around 0.06\mJyPerBeam, but we did not detect any significant structure near the pulsar. Then we generated images using \uv\ tapering with a larger FWHM to boost the signal to noise ratio (S/N). Tapering sizes are 20\arcs\ for the 3 and 6\cm\ images and 70\arcs\ for the 13 and 21\cm\ images. The 3, 6, and 13\cm\ images were produced with Brigg's robust parameter of 0.5 to suppress side lobes. For the 21cm image, we used natural weighting to maximize the sensitivity. 

For image decovolution, we first used the task \texttt{mossdi} to clean strong point sources in the Stokes I, Q, and U images. The residual maps were then cleaned simultaneously using \texttt{pmosmem} and the models were restored with beam sizes of 20\arcs\ in 3 and 6\cm\ and 70\arcs\ in 13 and 21\cm\ maps. The rms noise is around 0.06, 0.06, 0.7, and 0.8\mJyPerBeam\ in the Stokes I images and around 0.05, 0.04, 0.5, and 0.5\mJyPerBeam\ in the Stokes Q and U images at 3, 6, 13, and 21\cm, respectively. Finally, we generated polarization maps with the task \texttt{impol}. Besides, we have also applied the same procedure to produce full Stokes images of the pulsar using the on pulsed data.

\section{RESULTS}
\label{sec3}
\subsection{Morphology} 
\begin{figure*}
    \centering
    \includegraphics[scale=0.6]{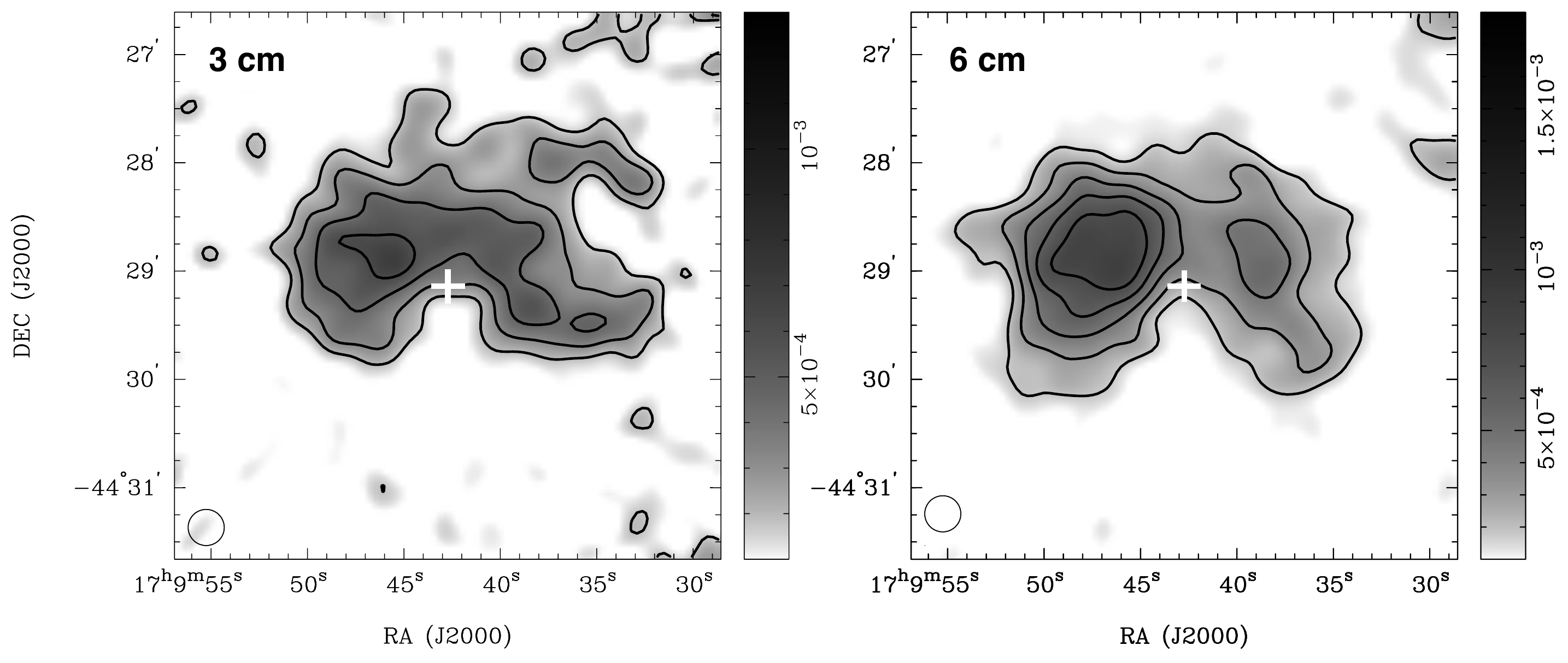}
    \caption{Total intensity images of the B1706 PWN at the 3 and 6\cm\ bands in the off-pulse phase with the pulsar emission excluded. The gray scale bar on the right is in units of\,\JyPerBeam. The crosses at the center of the images show the position of \psr. The circular beams at the bottom left indicate the beam size of FWHM 20\arcsec\ for both images. The rms noise in both bands is around 60\,\uJyPerBeam. The contours correspond to total intensity levels of 0.18, 0.3, 0.45, and 0.6\mJyPerBeam. }
    \label{CX_I}
\end{figure*}

Figure~\ref{CX_I} shows the total intensity maps of B1706 PWN during the off-pulse phase in the 3 and 6\cm\ bands. The radio PWN is clearly detected. It is elongated in the east-west direction with a size of $\sim4\arcm\times2\arcm\ $and wraps \psr\ in the north. The eastern part of \pwn\ is generally brighter, and the flux density peaks at 0.7\arcm\ and 0.9\arcm\ east of the pulsar, reaching 0.60 and 0.85\mJyPerBeam\ at 3 and 6\cm, respectively. At 3\cm, the nebula has a more uniform brightness distribution than at 6\cm, and it shows arc-like structure overall.
We also found a few protrusions in the PWN, one extends 2\arcm\ north of the pulsar and two others northwest and southwest from the pulsar extend towards west 2\arcm\ away from the pulsar.     
All these protrusions are not thicker than 0.5\arcm.
The protrusion features should correspond to data with \uv\ coverages from $\sim4.5$ to $\sim15$\,k$\lambda$, which are included in the 3\cm\ data. 
These features are therefore less likely to result from the missing flux problem. 
We need more observations to confirm this feature.
At 6\cm, the PWN shows two distinct peaks, resembling two lobes bracketing the pulsar.
The eastern part has an elliptical shape of 2\arcmin\ in size.
It is brighter than the western part and contains 67\% of the flux density of the PWN.
The western lobe is fainter and more elongated.
It has a size of $2\arcmin\times 1.5\arcmin$ and is oriented along the northeast-southwest direction.
Its surface brightness peaks at $\sim 0.8\arcm$ west of the pulsar and is only about 2/3 of that of the peak in the eastern lobe.
For both radio images, there is a ``bay" feature south of the pulsar with no detectable radio emission. The 3$\sigma$ flux density limit is around 0.18\mJyPerBeam\ in both the 3 and 6\cm.
The pulsar emission is clearly detected in the on-pulse data in both 3 and 6\cm\ bands with flux densities of 1.0$\pm$0.1\mJyPerBeam\ and 2.9$\pm$0.1\mJyPerBeam, respectively. 

\begin{figure}[hp!]
    \centering
    \includegraphics[scale=0.68]{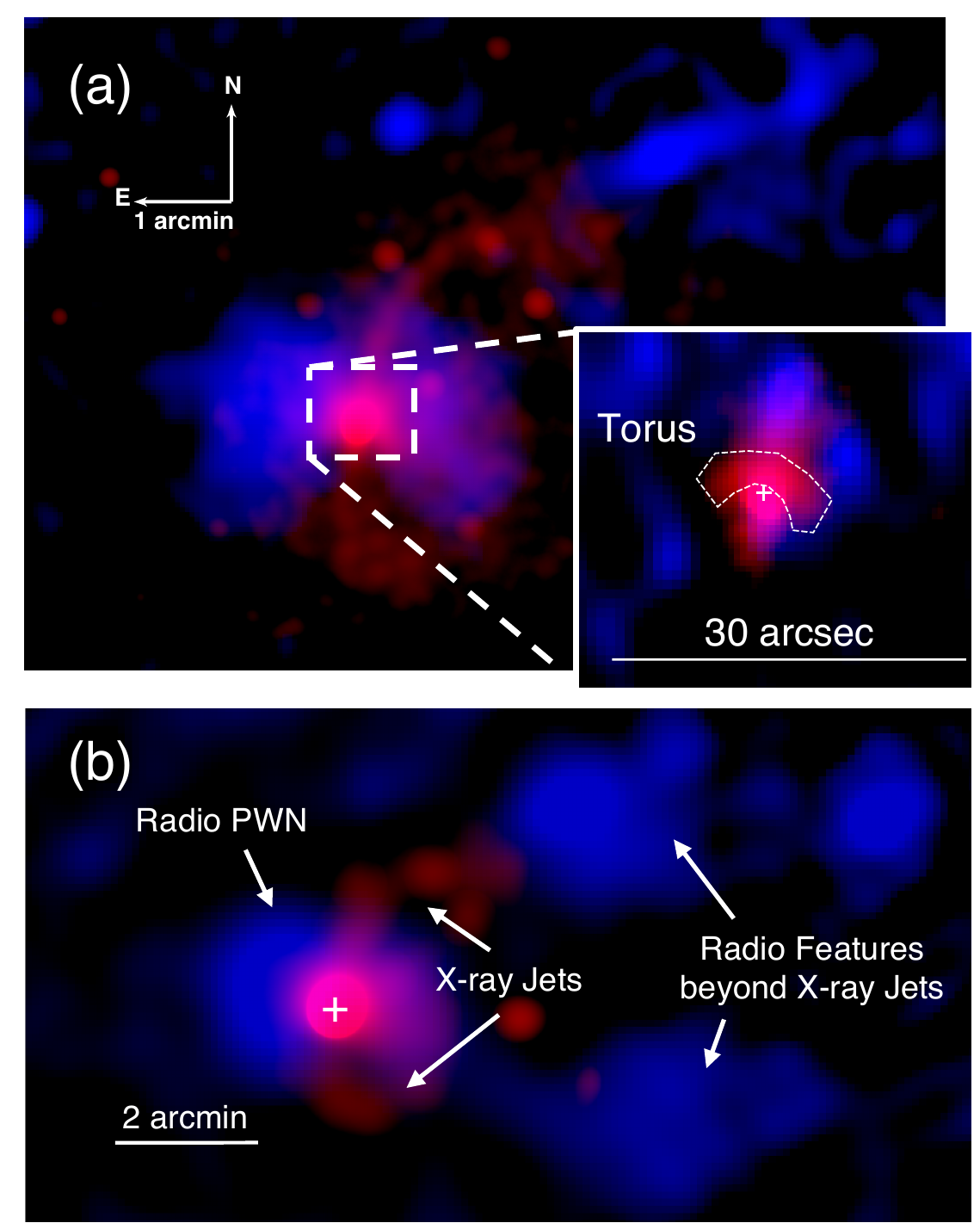} 
    \caption{(a): Comparison between radio and X-ray emission of B1706 PWN. The 6\cm\ radio emission is shown in blue with the 20\arcsec\ restored beam. The \emph{Chandra} 0.5--7 keV X-ray image is shown in red, also smoothed to 20\arcsec\ resolution. The inset image is the comparison of the X-ray torus/jet feature and the 3\cm\ high resolution radio image, with the X-ray torus highlighted by the white region}. (b): Same as (a) but both  are smoothed to 50\arcsec\ resolution. The cross in white indicates \psr. 
    \label{X_R_RGB}
\end{figure}

In Figure~\ref{X_R_RGB}, we compare the radio images with a 0.5--7\,keV X-ray image obtained with the \emph{Chandra} X-ray Observatory. We compared the X-ray torus/jet feature close to the pulsar with the 3\cm\ radio image with a beam of 6.3\arcs$\times$3.5\arcs, and found no counterpart of the X-ray PWN. The 3\cm\ radio image has a rms noise of 0.02\mJyPerBeam. We also smoothed the X-ray image to 20\arcsec, same as that of the radio image. Similarly, there is X-ray emission but no radio emission in the inner PWN, and the radio emission only appears in the outer PWN beyond 10\arcsec\ from the pulsar.
Meanwhile, the X-ray emission fades away in the outer PWN $\sim$25\arcsec\ from the pulsar.
The 3 and 6\cm\ images also show a linear, jet-like structure extending west from the end of the northern X-ray jet with a flux density of 8.2$\pm$0.6\,mJy at 6\cm.
It has a length of $\sim$3\arcmin\ and the width is not clearly resolved by the 6\,cm observation (see Figure~\ref{X_R_RGB}a).
We also find similar emission beyond the southern X-ray jet after smoothing the 6\,cm intensity map to 50\arcsec\ (see Figure~\ref{X_R_RGB}b) but it is fainter and more diffused than the emission in the north. 
More data are needed to confirm these features.

Figure~\ref{LS_I} shows the total intensity images of the overall SNR in the 13 and 21\cm\ bands. This is the first time that the 13\cm\ image is shown, and the SNR shows a similar morphology to that at 21\cm:
there is a $\sim$40\arcm\ semicircular rim in the west and a bright east-west ridge in the south $\sim$30\arcm\ connecting the pulsar to the western rim. \psr\ is located at the tip of the ridge rather than at the center of the SNR.
The pulsar emission is visible in the images, since no pulsar binning mode was used in these observations. There is also significant emission detected at the locations of radio outer jets at 6\cm.
\begin{figure*}[tp]
    \centering
    \includegraphics[scale=0.62]{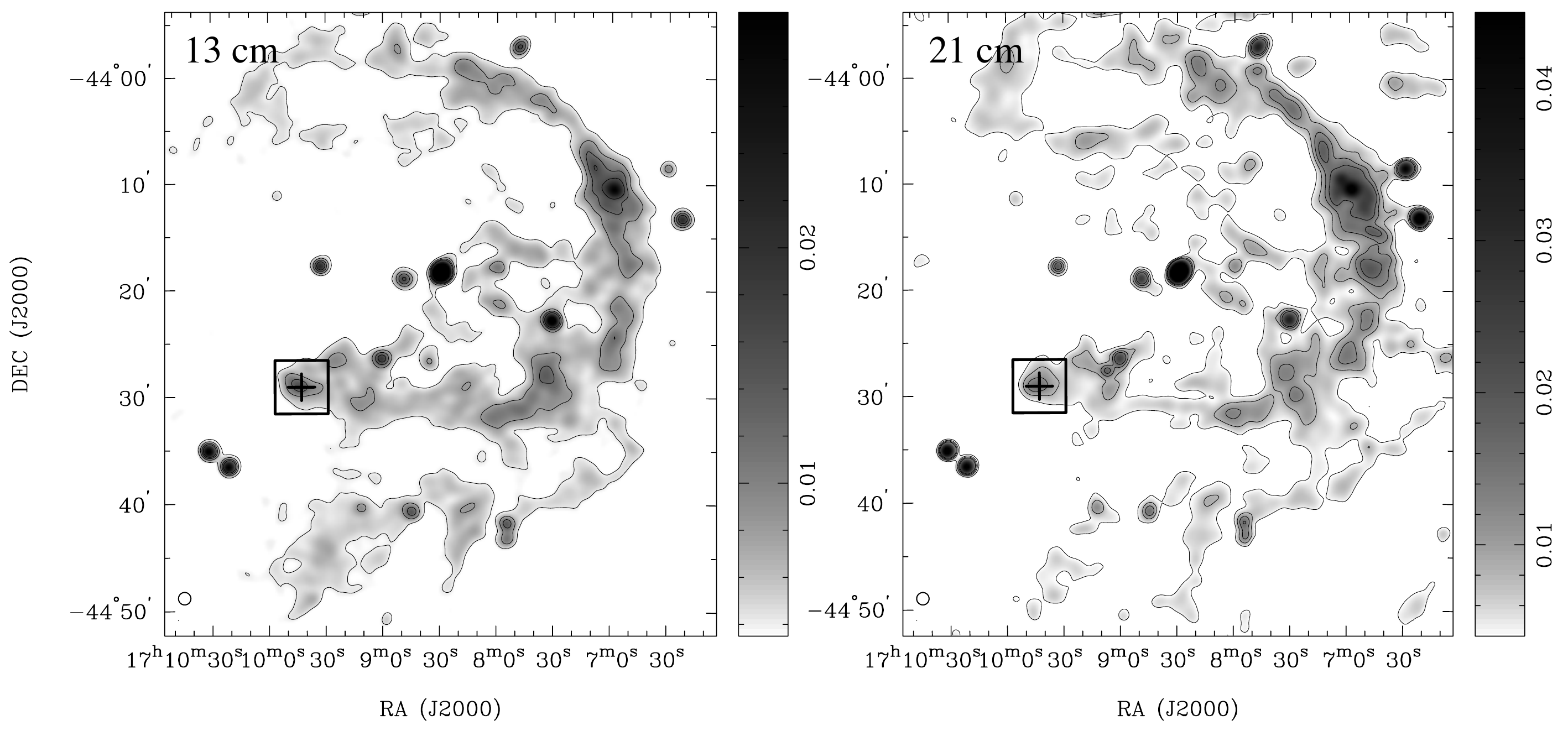}
    \caption{Total intensity maps of SNR \gname\ in 13\cm\ and 21\cm\ bands. The contours are at levels of 4, 8, 12, and 16\mJyPerBeam. The gray scale bars on the right have units of\,\JyPerBeam. The boxes indicate the field of view of Figure 1. Both images have a beam size of FWHM 70\arcsec, which is shown in the bottom left. The rms noise at 13\cm\ is around 0.7\mJyPerBeam and at 21\cm\ is around 0.8\mJyPerBeam. }
   \label{LS_I}
\end{figure*}

\subsection{Radio spectrum}
\begin{deluxetable}{ccccc}[b]
    \ \\
\tablenum{2}
\tablecaption{Flux density of the pulsar, entire PWN, and different components \label{tab:2}}
\tablehead{ \colhead{Freq.} & \colhead{Total} & \colhead{Eastern} & \colhead{Western} & \colhead{Pulsar} \\
\colhead{Bands} & \colhead{PWN (mJy)}  & \colhead{Lobe (mJy)}  & \colhead{Lobe (mJy)} & \colhead{(mJy)} }
\startdata
3\,cm & 18.5$\pm$1.0 & 8.9$\pm$1.0 &10.5$\pm$0.8 &1.0$\pm$0.1\\
6\,cm & 21.2$\pm$0.7 &14.1$\pm$0.8 &7.3$\pm$0.4 &2.9$\pm$0.1 \\
13\,cm& 39.2$\pm$1.4 & --        & --      &8.0$\pm$0.1 \\
21\,cm& 47.4$\pm$4.0 & --        & --      &10.2$\pm$0.4 \\ 
\enddata 

\end{deluxetable}

\begin{figure}[tp]
    \centering
    \includegraphics[scale=0.33]{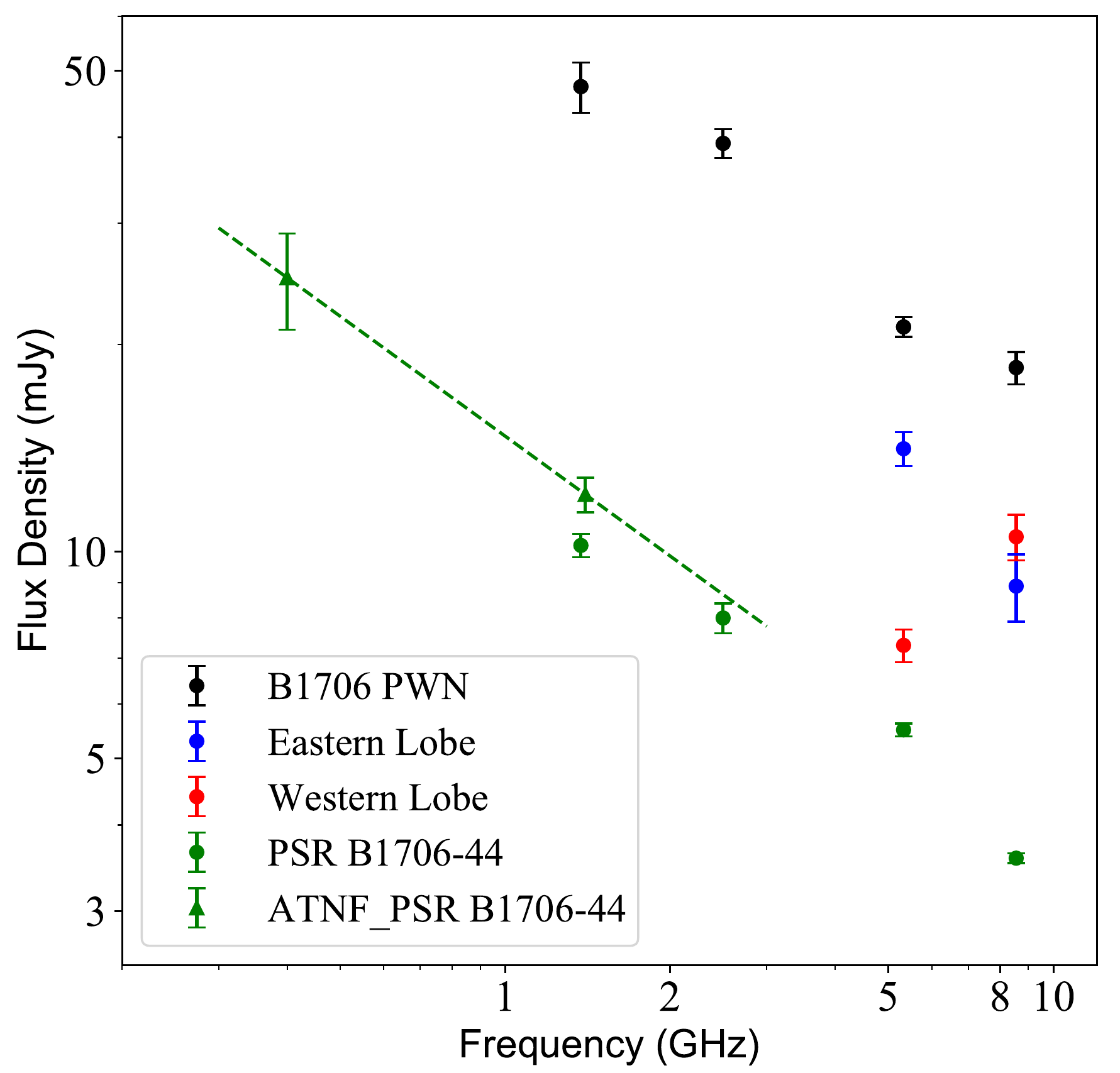}
    \includegraphics[scale=0.33]{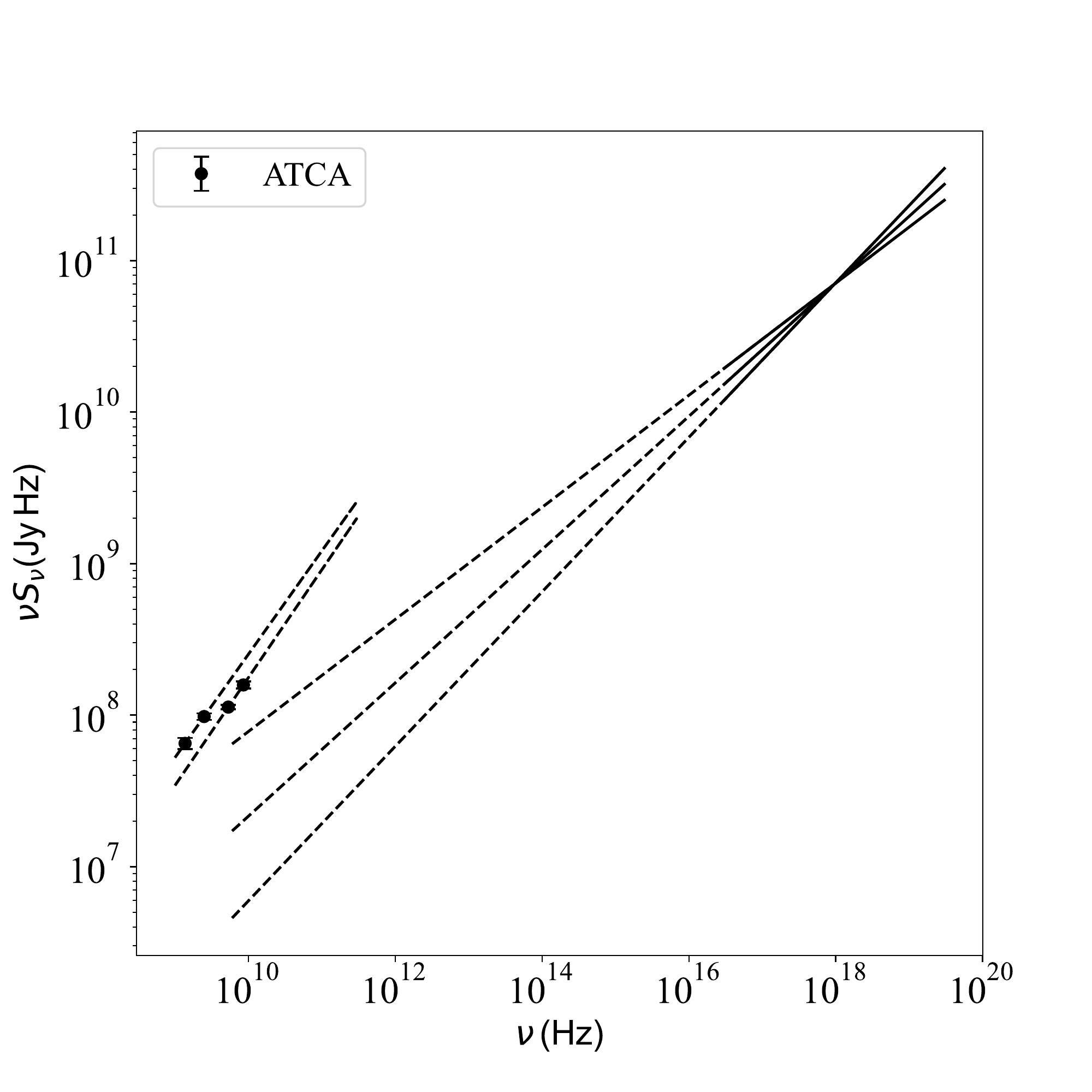}
    \includegraphics[scale=0.44]{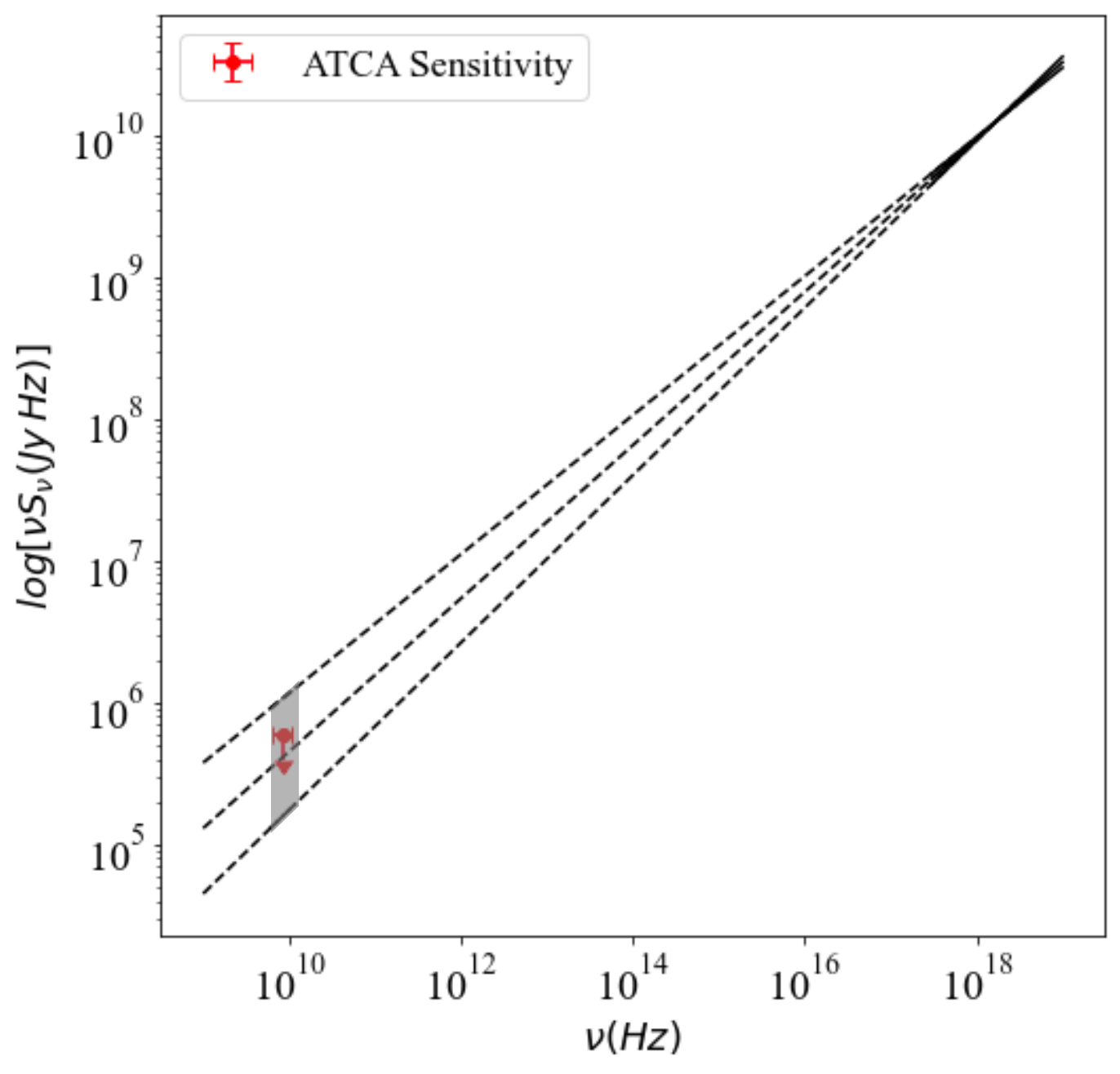} 
    \caption{Top: radio spectra of the overall B1706 PWN and different components. The green dashed line shows the extrapolated spectrum of the pulsar emission from 0.4 and 1.4\,GHz data. The flux densities of \psr\ at 0.4 and 1.4\,GHz are from the ATNF pulsar Catalog \citep{Manchester2005psrcatalog} and are shown as triangles with error bars. Middle: SED of the PWN from radio to X-ray bands. The black dots with error bar represent the measurement obtained with ATCA, and the lines in the top right show the best-fit unabsorbed X-ray spectrum obtained from \emph{Chandra}. Bottom: Multiwavelength SED of the X-ray torus. The X-ray spectrum is extrapolated to 3\cm\ wavelength (the band in gray), and the upper limit in red shows the 3$\sigma$ rms noise of the 3\cm\ observations. }
    \label{Spec}
\end{figure}

We measured the flux densities of the overall B1706 PWN and each component in different bands.
Background subtraction was performed using measurements from nearby source free regions.
The regions selected to measure the whole PWN and the eastern and western lobes in both bands are shown in Figure \ref{fig:SNR_regions}.

The estimated flux densities of the entire PWN are 18.5$\pm$1.0, 21.2$\pm$0.7, 39.2$\pm$1.4, and 47.4$\pm$4.0\,mJy in 3\cm, 6\cm, 13\cm, and 21\cm, respectively.
All these are plotted in Figure~\ref{Spec} and shown in Table~\ref{tab:2}.
The values at 13 and 21\cm\ have been subtracted for the pulsar flux density and those at 3 and 6\cm\ are measured from the off-pulse phase images. 
Due to the lack of short \uv\ spacing below 0.8\,k$\lambda$ in the 3 and 6\cm\ observations, the maps have low sensitivity to structures larger than $4\arcm$.
In this case, we separately derived the spectrum from the two higher and lower frequency bands.
They both give a similar spectral index $\alpha \approx-0.3$ $(S_\nu\propto\nu^\alpha)$  (see Figure~\ref{Spec}).
We note that our flux density measurement of the overall PWN at 6\cm\ ($\sim$21\,mJy) is comparable to the one measured with the VLA ($\sim$28\,mJy), although slightly lower \citep{giacani_2001_psrB1706}. 
As the VLA data has similar \uv\ coverage as our ATCA observations.
The discrepancy could be due to different choices of source and background regions.
The flux densities of the eastern lobe are 14.1$\pm$0.8\,mJy and 8.9$\pm$1.0\,mJy in 6 and 3\cm, respectively; and those of the western lobe are 7.3$\pm$0.4\,mJy and 10.5$\pm$0.8\,mJy, respectively.
These give $\alpha=-0.97$ for the eastern lobe and $\alpha= +0.78$ for the western lobe.

We also plotted the 3 and 6\cm\ images after filtering data to the same \uv\ coverages, so that data in both bands have the same missing flux problem for a correction of the spectral index.
We obtained a spectral index $\alpha\sim 0$ for the whole PWN and spectral indices of $-0.05$ and $+0.96$ in the eastern lobe and the western lobe, respectively.
The latter is rather unusual and more observations are needed to confirm this. 

We also estimated the pulsar flux densities with extraction region same as the beam size. 
For measurements at 3 and 6\cm, we generated images with only on-pulse bins to show the pulsar emission. 
The pulsar has flux densities of 1.0$\pm$0.1, 2.9$\pm$0.1, 8.0$\pm$0.1, and 10.2$\pm$0.4\,mJy at 3, 6, 13, and 21\cm, respectively. 
The results are plotted in Figure~\ref{Spec}. 
Due to low resolution, the measurement at 21\cm\ could be contaminated by the PWN emission, but we note that the result is consistent with the pulsed flux density obtained from the single dish Parkes Radio Telescope, and the flux densities in all bands are in line with the extrapolation of the pulsar spectrum \citep{1706.4Gflux,10.1093/mnras/stx2476}.

We performed a multiwavelength comparison with the \emph{Chandra} X-ray data. We reprocessed all archival data of B1706 PWN using the CIAO software package \citep{Fruscione2006CIAOchandra}, then extracted the X-ray PWN spectrum using \texttt{specextract} to fit the background subtracted spectrum with an absorbed power law.  We obtained a photon index $\Gamma \sim 1.53\pm0.07$ and a total unabsorbed flux $f_{pwn}$=12.2$\pm$0.1$\times$ 10$^{-13}$\, erg\ cm$^{-2}$\ s$^{-1}$ in the 0.5--7\,keV range (excluding the pulsar emission). This gives $\alpha_{X}=-0.53$ in the X-rays. We plot the spectral energy distribution (SED) of the PWN from radio to X-ray bands in Figure~\ref{Spec}. A comparison with $\alpha_{radio}\sim-0.3$ in the radio band suggests a spectral break of $\Delta\alpha\sim0.2$. However, we note that the extrapolation of the radio and X-ray spectra do not intersect. This could be due to the X-ray observations being insensitive to faint emission in the outer PWN region. In this case, emission from the inner PWN region would dominate and the obtained photon index would be smaller than that of the overall PWN.

We also did such a comparison in high resolution about the X-ray torus region. For the X-ray torus, recent \emph{Chandra} results show a flux $f_{torus}$=1.26$\pm$0.03\ergcmPerSecond\ from 0.5 to 7\,keV with a photon index $\Gamma$=1.46$\pm$0.05 \citep{deVries_J1709_motion_2021}. The flux density in radio is estimated in a region shown in Figure \ref{X_R_RGB} and show a sensitivity of 0.06\,mJy for the torus region. We compared the radio flux density with the extrapolated X-ray spectrum. The SED is shown in Figure \ref{Spec}. 
\begin{figure*}[]
    \centering
    \includegraphics[scale=0.9]{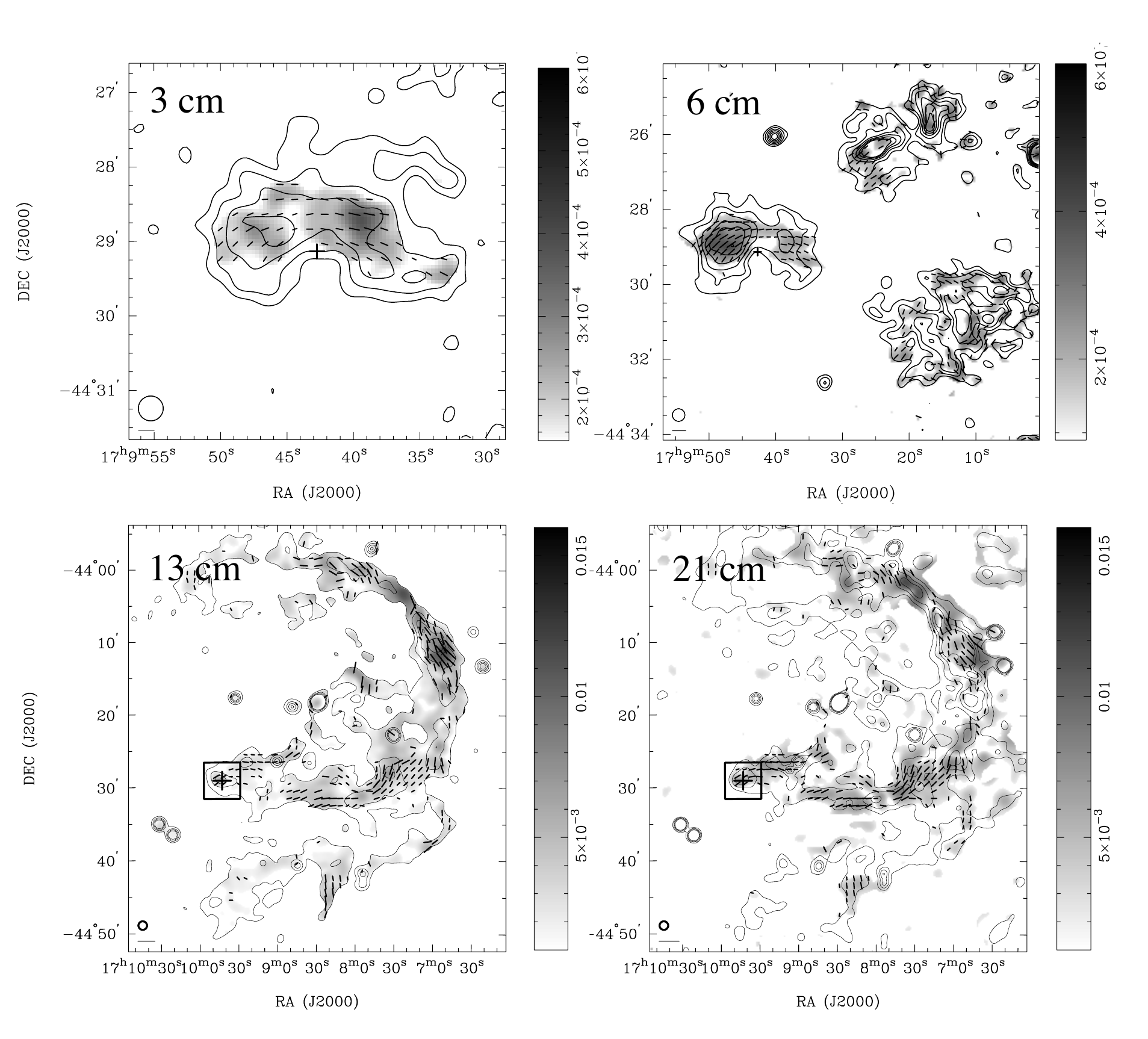}
    \caption{Linear polarized intensity maps of B1706 PWN and SNR \gname, overlaid with the total intensity contours and polarization vectors that show the intrinsic $B$-field orientation. The contours are at levels of 0.18, 0.3, 0.45, and 0.6\uJyPerBeam\ for 3 and 6\cm, and 4, 8, 12, and 16\mJyPerBeam\ for 13 and 21\cm. The gray scale bars correspond to the polarized intensity level in units of \JyPerBeam. The vector length is proportional to the polarization intensity, with the bars in lower left indicating polarization intensity of 0.5\mJyPerBeam\ in 3 and 6\cm\ maps and 10\mJyPerBeam\ in 13 and 21\cm\ maps. The black cross represents the position of \psr. As shown at the bottom left, the restoring beam sizes are 20\arcsec\ FWHM for 3 and 6\cm\ images and 70\arcsec\ FWHM for 13 and 21\cm\ images.} 
   \label{LSCX_PL} 
\end{figure*}

\subsection{Polarization} 
Figure~\ref{LSCX_PL} shows the polarized emission of the PWN and the SNR. We clipped the 3 and 13\cm\ maps where the polarization intensity has a signal-to-noise ratio (S/N) $<$3, total intensity S/N $<$5, or uncertainty of the position angle (PA) $>$10$^\circ$.
For the 6 and 21\cm\ maps, we applied the same clipping criteria for the polarization intensity S/N and PA, but $<$3 for the total intensity S/N.
The PWN is highly linearly polarized in all the bands and the polarized emission generally follows the total intensity.
The 3\cm\ polarization emission is east-west elongated and the size is $\sim4\arcm \times 1\arcm$.
However, it shows a peak $\sim0.8\arcm$ west of the pulsar, different from that of the total intensity map.
Whereas, the 6\cm\ polarization image shows two-lobed structure 
resembling the total intensity emission.
The eastern lobe is brighter and has a peak flux density of 0.37\mJyPerBeam. 
Polarized emission was detected in both the northern and southern radio features beyond X-ray jets, but the point source in northern one is unpolarized.
The linear polarization fraction in both 3 and 6\cm\ bands is around 45\% for the entire PWN, and around 85\% for the pulsar.
The circular polarization fraction of the pulsar is around 15\%
in both bands.  

In the 13 and 21\cm\ images, we found polarized emission on large scales including the PWN, the ridge, and the SNR shell. The PWN is detected at the end of the ridge with a blob-like structure aligning with the total intensity contours. The PWN polarized emission is fainter than that of the ridge and the SNR shell. The polarization fractions of the PWN in both 13 and 21\cm\ are around 30\%. The shell structure of the SNR shows significant polarized emission. The polarization fractions of the entire SNR are around 50\% and 40\% at 13\cm\ and 21\cm, respectively. 

\subsection{Rotation measure and intrinsic magnetic field orientation} 
The observed PA of the polarization vectors are rotated due to Faraday effect in the interstellar medium. The amount of rotation is proportional to the rotation measure (RM) times square of the wavelength ($\lambda^2$). We attempted to derive a high resolution RM map using the 3 and 6\cm\ data, but it has too large uncertainty to be useful. Therefore, we simply used the RM of the pulsar \citep[0.7$\pm$0.07\RMu;][]{Johnston-psrB1706-RM} to derotate the polarization vectors at these two bands. To determine the RM of the SNR, we selected edge channels with 32\,MHz bandwidth from the 13 and 21\cm\ data to generate Stokes Q and U maps. We used a \uv\ taper of 80\arcsec\ FWHM to boost the S/N. The images were then deconvolved using the same procedure as mentioned above, and restored with a circular beam of FWHM 80\arcsec, which is the resolution of the lowest frequency band. We formed four PA maps and applied a linear fit to determine the RM value at each pixel. The result is plotted in Figure~\ref{fig:RM} and the typical uncertainty of the map is $\sim$1\,rad\,m$^{-2}$. We found that the RM of the SNR varies from $-90$\RMu\ to +92\RMu\ and it is $\sim 0$\RMu\ near the pulsar position. The latter is in line with the RM of \psr\ \citep{Johnston-psrB1706-RM}. The RM of the PWN and the ridge varies smoothly compared with that in the SNR rim. 
\begin{figure*}
    \centering
    \includegraphics[scale=0.75]{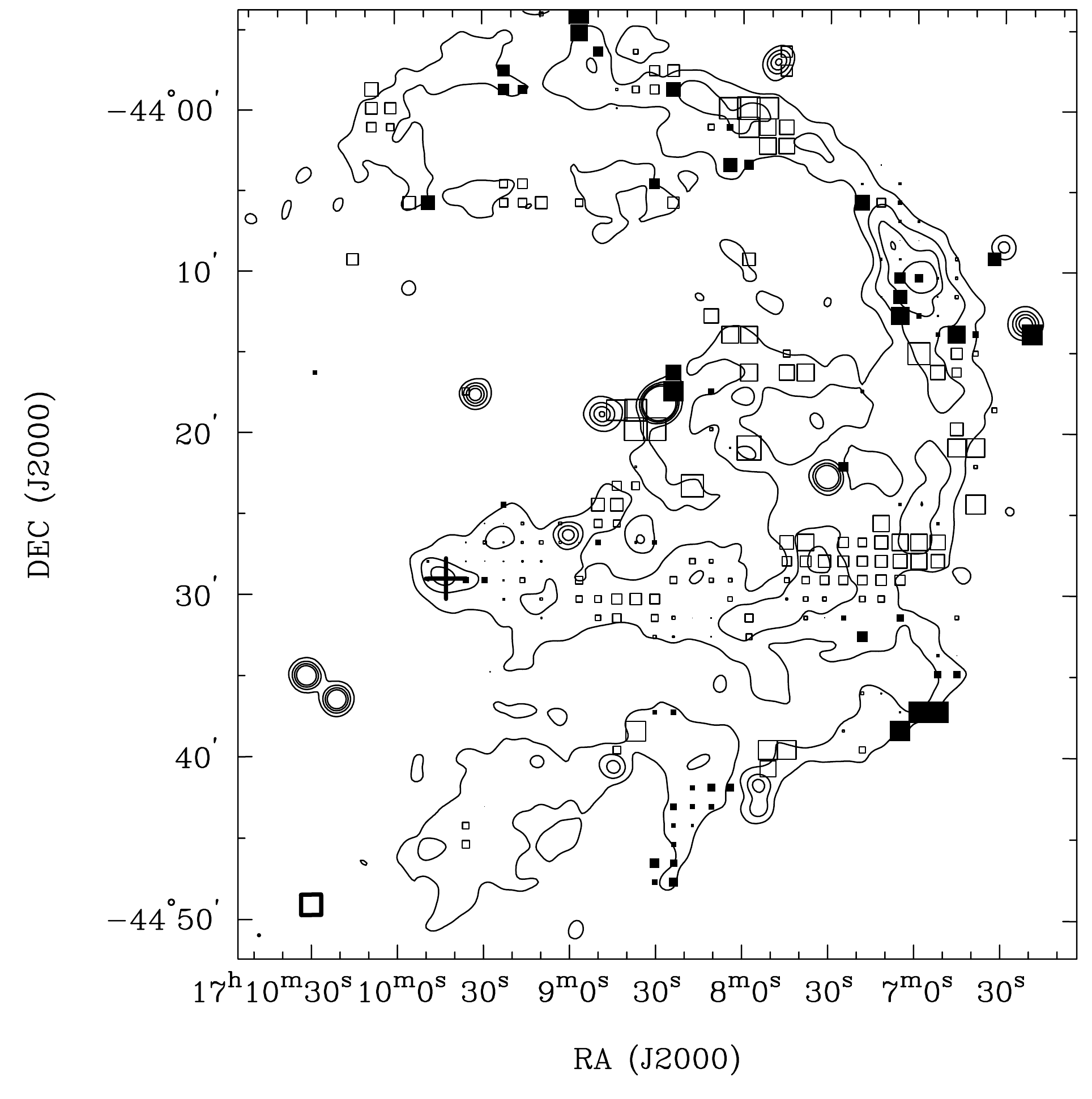}
    \caption{RM map of SNR \gname. The RM values vary from $-$90\RMu\ to $+$92\RMu, with the solid squares representing positive values and hollow boxes for negative. The contours are total intensity levels at 4, 8, 12, and 16\mJyPerBeam\ at the 13\cm. The cross indicates the position of \psr\ and the box at bottom left corresponds to RM value of $-50$\RMu.}
    \label{fig:RM}
\end{figure*}

Figure~\ref{LSCX_PL} shows the intrinsic magnetic field direction of B1706 PWN at 3 and 6\cm\ and of the SNR at 13 and 21\cm\ after correcting for the Faraday effect. The magnetic field of the PWN is highly ordered. It is oriented along the PWN elongation and wraps around the pulsar in the north, indicating a toroidal configuration.
For the outer jets, only faint polarized emission is detected, we are therefore only able to determine the polarization angle in the brightest regions.
At large scale, the magnetic field of the ridge well aligns with its elongation.
It then gradually switches to tangential along the rim of the SNR shell.

\section{DISCUSSION}
\label{sec4}
\begin{figure*}[ht!]
    \centering
    \includegraphics[scale=0.9]{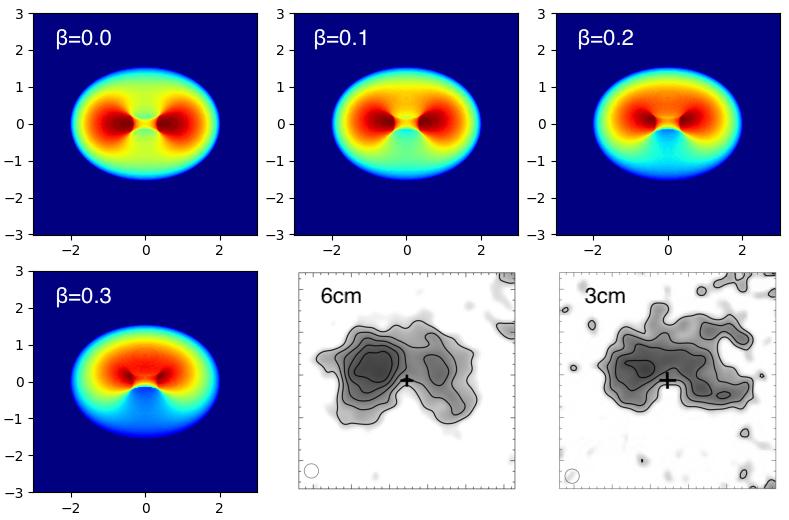}
    \caption{Models of a thick torus with Doppler boosting effect with different $\beta$ (from 0.0 to 0.3) in the case of B1706 PWN. The total intensity images at 6 and 3\cm\ are shown at the bottom panels for comparison.}
    \label{fig:Torus+Doppler}
\end{figure*}

\subsection{PWN structure}
Our radio intensity maps of the B1706 PWN reveal an overall arc-like morphology.
The emission is bright in the outer region of PWN but faint in the inner part, which is in contrast to the X-ray emission.
Similar X-ray--radio anti-correlation is also found in a few other PWNe, including Vela PWN, DA~495, G76.9$+$1.0, G319.9$-$0.7, G327.1$-$1.1 \citep{Dodson2003VelaPWN,Kothes2008DA495,Arzoumanian2011g76.9,Kargaltsev2008_J1509J1740,Ng2010_J1509g319.1,Ma_snailPWN_2016}. 
The cause is not clearly understood.
It was suggested that the radio emission in the inner PWN could be too faint to detect.
As the outflow decelerates, the particle number density increases outward, resulting in brighter radio emission.
On the other hand, synchrotron cooling makes the X-ray emission invisible in the outer PWN \citep{Kargaltsev2008_J1509J1740}.
We applied this idea to B1706: we extrapolated the X-ray spectrum of the torus reported by \citet{deVries_J1709_motion_2021} down to the radio band with a simple unbroken power law.
This suggests a flux density from $\sim$0.01 to $\sim$0.2\,mJy  at 3\cm. From our highest resolution 3\cm\ intensity map, the  3$\sigma$ limit at this region is $\sim$0.06\mJyPerBeam, giving a detection limit of $\sim$0.06\,mJy based on the torus area (see Figure \ref{Spec}).  
In this case, we still cannot firmly rule out the scenario that the X-ray torus follows a simple unbroken power-law distribution from radio to X-rays. A further non-detection with about 3 times better sensitivity is required to reject such a scenario. Alternatively, we note that the injected spectrum could have an intrinsic spectral break, if the particle acceleration is due to magnetic reconnection in the termination shock or Weibel instability \citep{Lyubarsky2001magnet_recon_stripewind,Weibel1959Weibelinstability}. In this case, the radio emission could be a few orders of magnitude fainter, making it very difficult to detect. Deeper radio observations are needed to discriminate between these.

Our high resolution radio maps reveal overall arc-like structure for B1706 PWN,
which is similar to Vela, Boomerang (G106.6+2.9),
and G76.9+1.0
\citep{Dodson2003VelaPWN,Kothes2006Boomerang, Arzoumanian2011g76.9}.
It was suggested that this could be caused by the passage of supernova reverse shock
and a thick toroidal model has been developed \citep{Chev11tori}. 
We tried to apply the same model to our 3\,cm image, but found that it fails to explain the lack of emission in the bay south of the pulsar.
Indeed the model is always symmetric and also cannot explain the tongue-like morphology of the Boomerang.
It is worthy mentioning that many X-ray PWNe also show an asymmetric torus feature close to the pulsar due to the Doppler boosting effect, such that emission is brighter if the particles are moving toward the observer and vice verse.
We therefore added Doppler boosting effect to the model following a similar procedure as  \citet{PWNTori_Ng2004,Ng2008PWNTori}.
We built a torus in 3D with circular cross-section and an outer radius of 2\arcm, assuming uniform and isotropic emission inside. We set the viewing angle between the torus axis and the line of sight to be 53.3$^\circ$ \citep{2005ApJ...631..480R} and the outer boundary radius 6 times of the inner boundary radius. We considered Doppler boosting effect following  \citet{Pelling1987PWNTori}. The apparent intensity $I$ is 
\begin{equation}
   I\propto(1-n\cdot\beta)^{-(1-\Gamma)}I_{0},  
   \label{eq:Doppler_boosting}
\end{equation}
where $n$ is the unit vector from the observer, $\beta=v/c$ is the assumed velocity of the radial post-shock bulk flow, $\Gamma$ is the photon index in the rest frame, and $I_0$ is the intrinsic intensity of synchrotron emission taken to be constant.
We projected the model onto the plane of the sky to generate a 2D brightness map for comparison with the data. 
Figure~\ref{fig:Torus+Doppler} shows models of a radio B1706 PWN having a thick equatorial torus and the Doppler boosting effect with different $\beta$ values in the bulk flow, as well as the 6 and 3\cm\ radio images of the B1706 PWN.
In the scenario of $\beta=0$ (i.e., negligible Doppler boosting), the model shows two equatorial lobes both having a brightness peak inside, and a fainter region between the lobes.
We note that this model is not only horizontally symmetric, but also vertically symmetric.
Considering Doppler boosting will result in enhanced brightness in an approaching bulk flow and reduced brightness in flows leaving away, such that the upper part is brighter than the lower part in our model.
The brightness of the upper part becomes comparable to or even overwhelms that of the lobes in the scenario of $\beta\geq0.2$. 
In the latter case, the model has a kidney-shape feature wrapping the pulsar with a single peak north of the pulsar region.
We suggest that the model can capture characteristic features of the radio PWN observed, including the overall arc-like PWN wrapping the pulsar in the north and the faint bay in the south.
Comparisons between these models and the observations show that a constant value $\beta\sim0.2$ over the entire torus gives the best result of the 3\cm\ PWN, while the 6\cm\ PWN can be better described by $\beta\sim0.1$.

\begin{figure*}
    \centering
    \includegraphics[scale=0.72]{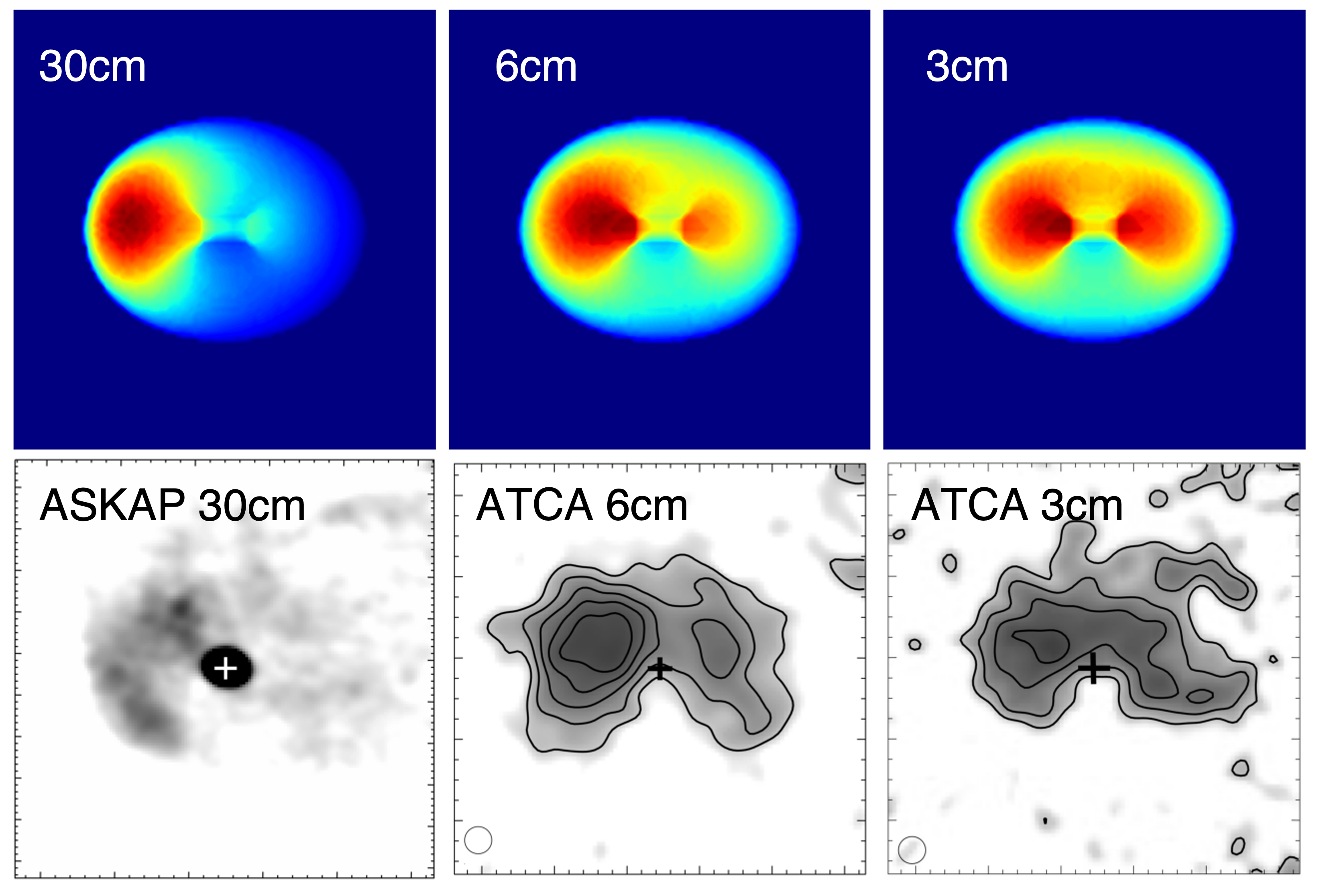}
    \caption{Models of the B1706 PWN in different wavelengths with a gradient in the east-west direction (\textit{the top row}), as well as the observation results of B1706 PWN at 30, 6, and 3\cm\ (\textit{the bottom row}) taken with ASKAP \citep{ASKAP2019AS101} and ATCA. 
    }
    \label{fig:thick_torus+doppler_boosting+alpha_gradient}

\end{figure*}

Our 6\cm\ image is slightly different from the 3\cm\ image (e.g., the gap between the eastern and western parts), so that the best fit $\beta$ are slightly different.  
To resolve this, we consider a spectral gradient across the PWN, as motivated by the different spectra between the eastern and western parts (see Figure \ref{Spec}). 
We fix $\beta=0.2$ and consider that $I_{0}$  depends on frequency as
\begin{equation}
    I_{0,\nu}=I_{0,\nu_{0}}\cdot( \frac{\nu}{\nu_0})^{\alpha},
\end{equation}
where $I_{0,\nu}$ is the intrinsic intensity in frequency $\nu$, and $I_{0,\nu_0}$ is the intensity in a reference frequency $\nu_0$. 
We also assume that $\nu_0=10\,\mbox{GHz}$ and 
\begin{equation}
    \alpha = 0.9\frac{x_{EW}}{R_{pwn}}, 
\end{equation}
where $x_{EW}$ is the position of a point in the east-west direction with a scale of the PWN radius $R_{pwn}$, indicating $\alpha=-0.9$ at the eastern end of the PWN and $\alpha=+0.9$ at the western end.
We then obtain the model at 3 and 6\cm\ and also extrapolate it to around 30\cm. 
Besides, we compare all these models with the 3 and 6\cm\ ATCA images and 30\cm\ ASKAP image \citep{ASKAP2019AS101}, and all these are shown in Figure \ref{fig:thick_torus+doppler_boosting+alpha_gradient}.  
Our simulation shows the PWN with a much brighter eastern part at 30\cm, a larger brighter eastern lobe and a smaller western lobe at 6\cm, and lobes connected from the north of the pulsar at 3\cm. 
These models can reproduce the main morphological features of the observed PWN in these bands. 
we also tried different values of $\beta$ but found that 0.2 gives the best fit result.
A direct comparison with $\beta_{X}=0.7$ in the X-ray emitting region \citep{2005ApJ...631..480R} suggests deceleration along the flow.
This also implies the particle accumulation, which could give rise to the observed anti-correlation between the radio and X-ray emission.

Our modeling result shows that B1706 PWN could have a toroidal structure in 3D, and it appears as arc-like due to Doppler effect. 
This model could be used to explain the arc-like morphology found in other radio PWNe, e.g., Vela and Boomerang. 
Finally, we note that toroidal structure can be resulted from a toroidal $B$-field as simulations suggest \citep{Porth2017torijetmodel}. 
This is supported by our polarization result, which reveals a toroidal $B$-field configuration.

\begin{figure*}
    \centering
    \includegraphics{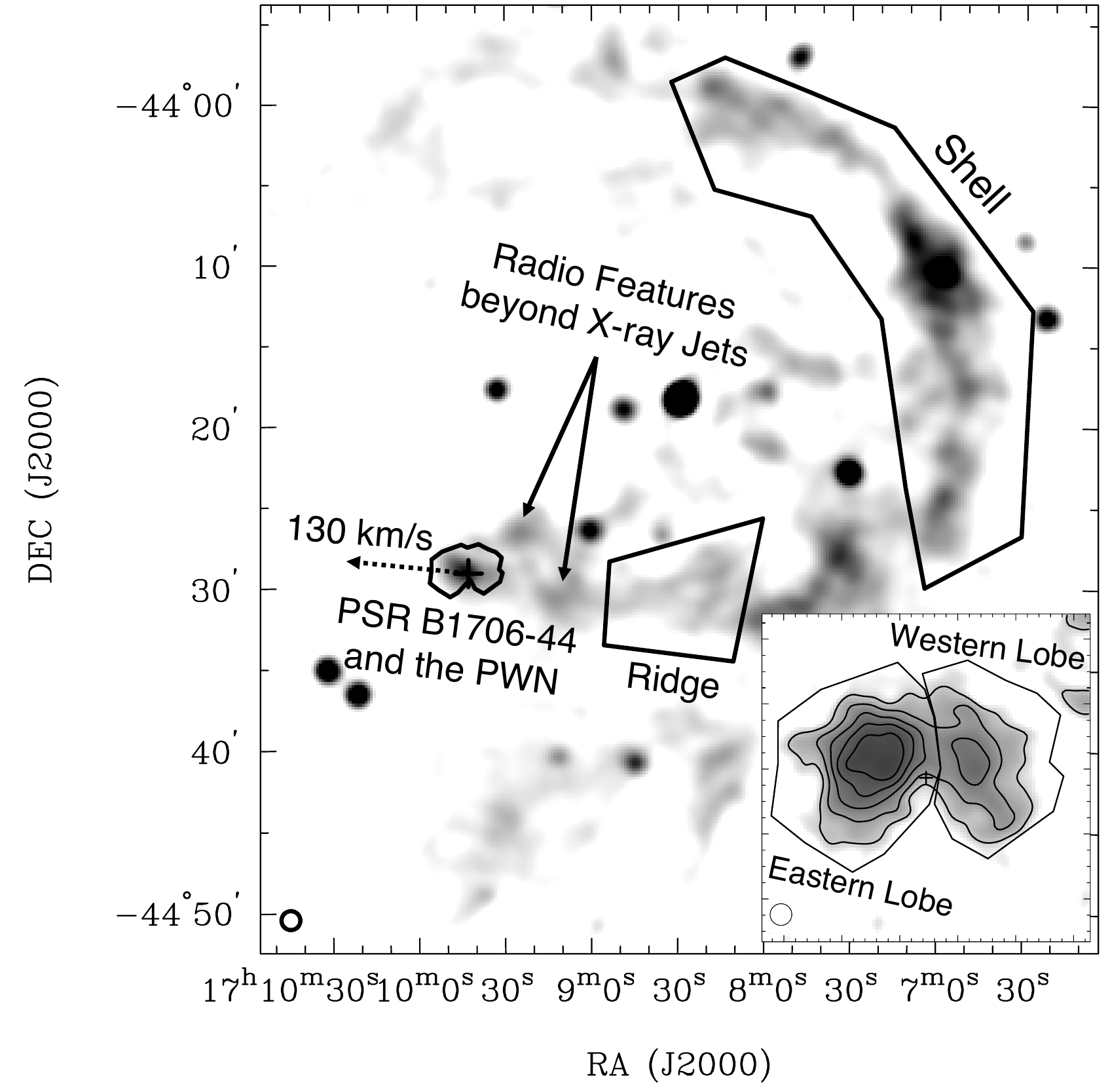}
    \caption{Total intensity image of SNR \gname\ at 13\cm. \emph{Inset:}
    Zoom in of the PWN at 6\cm. Extraction regions (e.g., shell, ridge, the whole PWN, and the eastern and western lobes) for spectral measurements are indicated. The circular beams of 20\arcsec\ (inset) and 70\arcsec\ (SNR) are shown bottom left of the images as scale references. The dashed vector shows the pulsar motion direction \citep{deVries_J1709_motion_2021}. \smallskip}
    \label{fig:SNR_regions}
\end{figure*}

\subsection{Equipartition magnetic field}
We estimated the equipartition $B$-field strength of the PWN 
\begin{equation}
B_{eq}=[6\pi(1+k)c_{12}L_{syn}\Phi^{-1}V_{pwn}^{-1}]^{2/7}, \label{Beq}
\end{equation}
where $V_{pwn}$ is the emission volume, $L_{syn}$ is the synchrotron luminosity, $\Phi$ is a filling factor for the emission (it is usually taken as 1 even though not 100\% of the volume emits), $k$ is the ratio between electron energy and the energy of heavy particles, $c_{12}$ is a constant related to synchrotron radiation process and weakly depends on the frequency range \citep{Pacholczyk1970B_equi_c12}.
We selected PWN flux density at 6\cm\ as a reference and assumed a simple power-law spectrum with a spectral index $\alpha\sim-0.3$ from 10$^{7}$ to 10$^{11}$\,Hz to obtain $L_{syn} = 7.9\times10^{30}d_{2.3}^2$\ergPerSecond, where $d_{2.3}$ is the source distance in units of 2.3\,kpc.
To estimate the volume of the PWN, we assumed an oblate spheroid for the emission volume in 3D.  The oblate spheroid is $ 2.6\arcm\times4.2\arcm\times4.2\arcm$ in size, which gives a volume of $V_{pwn}$=$2.2\times10^{56}$\cm$^3$. These give
\[B_{eq}=10.5(1+k)^{2/7}\Phi^{-2/7}d_{2.3}^{-2/7}\,\mu \mbox{G.}\]
Taking $k$=0 and $\Phi$=1, the $B$-field is in the order of 10\uG, slightly lower than 27\uG\ and 15\uG\ estimated for the inner and outer X-ray PWN, respectively \citep{deVries_J1709_motion_2021}. However, the decay in $B$-field strength is much slower than $B\propto 1/r$ predicted by theory \citep{Porth2017torijetmodel}, given that the radio PWN is $\sim$5 times larger than that in the X-rays.

We also roughly estimated the equapartition $B$-field of the linear radio feature beyond the X-ray jet northwest of the pulsar, assuming it associated with the PWN.
Following the same procedure as above and taking the emission volume as a cylinder,
the flux density measurements at 3 and 6\cm\ gives
\[B_{ls}=70(1+k)^{2/7}\Phi^{-2/7}d_{2.3}^{-2/7}\,\mu \mbox{G.}\]
The result is similar if we exclude the point source,
\[B_{ls^{\prime}}=59(1+k)^{2/7}\Phi^{-2/7}d_{2.3}^{-2/7}\,\mu \mbox{G.}\]
This is significantly higher than that of the main radio PWN, but is comparable to that of the X-ray bright inner PWN.
Identification the association between the PWN and this feature and requires more following observations and detailed modeling of the multiwavelength spectrum, which is beyond the scope of this work \citep[e.g.,][]{zhang2008NothermalPWN}.

\subsection{Nature of the ridge}
It is clear from our 13 and 21\cm\ maps that the ridge protruding east well aligns with the pulsar motion direction \citep{deVries_J1709_motion_2021}.
We therefore suggest that the ridge could be a pulsar tail instead of SNR structure.
In addition, our polarization maps show a good alignment between the $B$-field and the orientation of the ridge, which is a common feature of pulsar tails \citep[e.g., G319.9$-$0.7, G327.1$-$1.1, and the Mouse;][]{Ng2010_J1509g319.1,Ma_snailPWN_2016,Yusef-Zadeh2005mouse}. To confirm the nature of the ridge, we compared its radio spectrum with that of the SNR rim, using extraction regions shown in Figure~\ref{fig:SNR_regions}. We found a significantly flatter spectral index of $-$0.3 in the ridge than $-$1.1 in the shell, implying that the ridge likely comprises of pulsar wind. Moreover, the RM map shows a comparable RM as that of the pulsar and no significant variation in the entire ridge, therefore, supporting this idea. Although the ridge extends beyond the pulsar birth site suggested by \citet{deVries_J1709_motion_2021}, it could be formed by fast outflow or pulsar wind swept by the SNR reverse shock. The latter scenario could also explain the TeV emission \citep{G343.1-2.3TeV_HESS2011}. Deeper X-ray observations along the ridge can reveal any spectral evolution. Any enhanced synchrotron cooling could support the interaction with the supernova reverse shock. If confirmed, B1706 PWN could be in the same evolutionary state as Vela and Boomerang, both are suggested to be after the passage of reverse shock. We note that it remains unclear how the toroidal $B$-field in all these sources could be retained. It could be magnetic pressure against compression by the shock, or the radio PWNe are recently formed after the shock interaction. Further numerical simulations are needed to distinguish between these cases.
As is mentioned, the polarized emission of the ridge is brighter than that in the PWN. 
We suspect that the ridge region should be thicker than the B1706 PWN locating at the tip of it. 
It is also likely that the particles in the ridge have a larger density due to mixing with external materials (e.g., SNR ejecta) or particle slowing down.
Therefore, there are more particles emitting along the line of sight through the ridge than the PWN. 

\section{Conclusion}
\label{sec5}
We present a radio study of B1706 PWN using new and archival ATCA observations at 3, 6, 13, and 21\cm\ bands. Our main results are summarized below:

\begin{itemize}
\item  The 3 and 6\cm\ total intensity images show an arc-like structure with a scale of $\sim4\arcm\times2\arcm$ wrapping around \psr\ in the north. No radio emission is detected at the X-ray torus and jet location, and the radio PWN only brightens beyond 10\arcsec\ from the pulsar. We show that the radio PWN morphology can be fit by a thick torus model with Doppler boosting effect. The result suggests a bulk flow velocity of $0.2c$, lower than that in the X-ray torus.

\item  Our polarization results reveal a toroidal $B$-field for the PWN and we estimate a field strength of $\sim 10\mu$G assuming equipartition between particle and magnetic field energies.
This value suggests a slight decay compared with that of the X-ray bright region. 


\item  The ridge of the SNR has elongation and magnetic field well aligned with the pulsar proper motion direction.
It also has a radio spectrum flatter than the rest of the shell.
All these suggest that it could be a pulsar tail instead of a filamentary structure of the SNR.
\end{itemize}

\section*{ACKNOWLEDGMENTS}

We thank the anonymous referee for providing useful suggestions.
C.-Y. Ng is supported by a GRF grant of the Hong Kong Government under HKU 17301618. 
The Australia Telescope Compact Array is part of the Australia Telescope National Facility which is funded by the Commonwealth of Australia for operation as a National Facility managed by CSIRO.
This study also makes use of software in the application packages Miriad, CIAO and Sherpa.

\end{large}

\bibliography{references}

\end{document}